# Adding to the Regulator's Toolbox: Integration and Extension of Two Leading Market Models

Brian F. Tivnan, The MITRE Corporation
Matthew T. K. Koehler, The MITRE Corporation
Matthew McMahon, The MITRE Corporation
Matthew Olson, The MITRE Corporation
Neal J. Rothleder, The MITRE Corporation
Rajani R. Shenoy, The MITRE Corporation

**Abstract:**
As demonstrated during the recent financial crisis, regulators require additional analytical tools to assess systemic risk in the financial sector. This paper describes one such tool; namely a novel market modeling and analysis capability. Our model builds upon two leading market models: one which emphasizes market micro-structure and another which emphasizes an ecology of trading strategies. We address a limitation of market modeling, namely the consideration of only one dominant trading strategy (i.e., long positions). Our model aligns closely with several widely held stylized facts of financial markets. And a final contribution of this work stems from our empirical analysis of the fractal nature of both empirical markets and our market model.

Brian Tivnan
The MITRE Corporation
Farrell Hall, Office 216
210 Colchester Avenue
Burlington, VT 05405
USA
Phone: 802.656.2239
Fax: 802.656.5838
Electronic mail: btivnan@mitre.org

This manuscript was presented at the 2011 annual conference of the Eastern Economic Association.

With respect to the financial market modeling, the authors gratefully acknowledge: (1) several members of the MITRE technical staff for their help defining the models and producing some functional forms of the statistics and (2) additional support for this research from a Business Network Fellowship at the Santa Fe Institute. This paper is extended from "Agent-Directed Simulation for Systems Engineering: Applications to the Design of Venue Defense and the Oversight of Financial Markets" published in the *International Journal of Intelligent Control and Systems*, Vol. 14, No. 1, March 2009, 20-32.

## 1. INTRODUCTION

As demonstrated during the recent financial crisis, regulators require additional analytical tools to assess systemic risk in the financial sector. This paper describes one such tool; namely a novel market modeling and analysis capability. Our model builds upon two leading market models: one which emphasizes market microstructure and another which emphasizes an ecology of trading strategies. We address a limitation of market modeling, namely the consideration of only one dominant trading strategy (i.e., long positions). Our model aligns closely with several widely held stylized facts of financial markets. And a final contribution of this work stems from our empirical analysis of the fractal nature of both empirical markets and our market model.

This paper begins with an overview of market modeling and some of the leading models. We subsequently describe our attempts to replicate and integrate two of the leading models. We then describe a critical extension of the integrated model, that is the inclusion of margin and short positions. We conclude the paper with an analysis of the model validity and its comparison to empirical data.

## 2. MODELS OF FINANCIAL MARKETS

Models of financial markets can be particularly useful, allowing policy-makers to reason about changes to market regulation in a benign environment prior to implementation. Modeling of financial markets has been an objective of the analytical and practitioner communities alike for some time; this study builds on two leading models that are readily available in the academic literature.

### 2.1 Introduction

This study describes work to replicate, integrate and extend two agent-based models (ABM) of financial markets, that of Farmer and colleagues [1, 2] and that of Cont and colleagues [3]. This study builds on a modeling concept, namely, docking. The docking framework of Axtell and colleagues [4] is used herein to compare the replicated models to their original counterparts (i.e., referents) and demonstrate the relational equivalence between the referents and replicated models. This is a methodology important to ABMs as it can form a logical base for verification, validation, and accreditation of the ABMs used for policy analysis, decision-support, or even systems engineering. This study also highlights the importance of only making the ABM as complex as necessary. As will be demonstrated, very simple models can capture much of the dynamics exhibited by a system even one as complex as a financial market. As the topic of financial market modeling may be new to some readers this paper will begin with an overview of related agent-based market models, then it will describe the replication of the Farmer and Cont models. The work continues with a discussion of an extension to the Cont model and a hybrid model combining the market structure of Farmer's model with the trading agents of Cont's model. The paper will conclude with a discussion of the analytic side of this work; namely our replication of the Cont statistics and our use of a multi-fractal analysis to compare our ABM results with that of the S&P 500.

### 2.2 Overview of Agent-Based Models of Stock Markets

Agent-based models of equity stock markets began in the 1992-1993 timeframe with the Santa Fe Institute (SFI) market model [5]. Largely predicated on Holland's [6] genetic algorithms, the SFI market model was well received as a novel contribution, largely based on its qualitative agreement with empirical observations of market dynamics. Following the SFI market model, Lux and colleagues [7, 8] introduced a model with a single trade type (i.e., market orders) that was the first to demonstrate clustered volatility, one of the stylized facts common to many markets. A shortcoming of the Lux model is that its results proved to be sensitive to the size of the trader population. Similar to the Lux market with only market orders (defined, *infra*), LeBaron [9] introduced a market model with agents that learn based on a neural network. In the LeBaron model agents decided how much of their total wealth to invest; therefore successful agents can have a large impact in the market. Darley and colleagues [10] built a model of the NASDAQ market and were the first to infuse limit orders (defined, *infra*) into their model and they made five of six correct predictions about the NASDAQ transition to a decimal-based, tick size. Farmer and colleagues [1] built an empirically driven model of zero-intelligence traders intended to depict the structure of a continuous double auction. Cont and colleagues [3, 11, and 12] have developed a decentralized trader model that qualitatively represents the five prevailing stylized facts common to most modern markets and consistent over wide time periods. The traders in the Cont models have heterogeneous trading thresholds, and the traders—many of whom trade rather infrequently—adapt their thresholds based upon performance feedback.

### 2.3 Market Models

When modeling an artificial financial market the potential complexities are daunting. Further, as lamented in Ghoulmie et al [3] the addition of many of these features makes determining each feature's individual impact very difficult. This brings up a common problem with ABM design; namely, the simulation can become so complex that it is nearly as difficult to understand as the system it is designed to emulate. Therefore, some researchers have taken a different tack, specifically, how simple can a financial model be and still produce results that map to the real world in a non-trivial way? The work discussed here focuses initially on two of these models: the "Zero Intelligence model" as reported in Iori, et al. [2] and Farmer, et al. [1], and the "Cont model" as reported in Ghoulmie, et al. [3]. We found these two models of particular interest as they approach the problem from very different perspectives. The Farmer model stresses the importance of market structure; whereas the Cont model abstracts the market away almost entirely in favor of traders that have greater than zero intelligence.

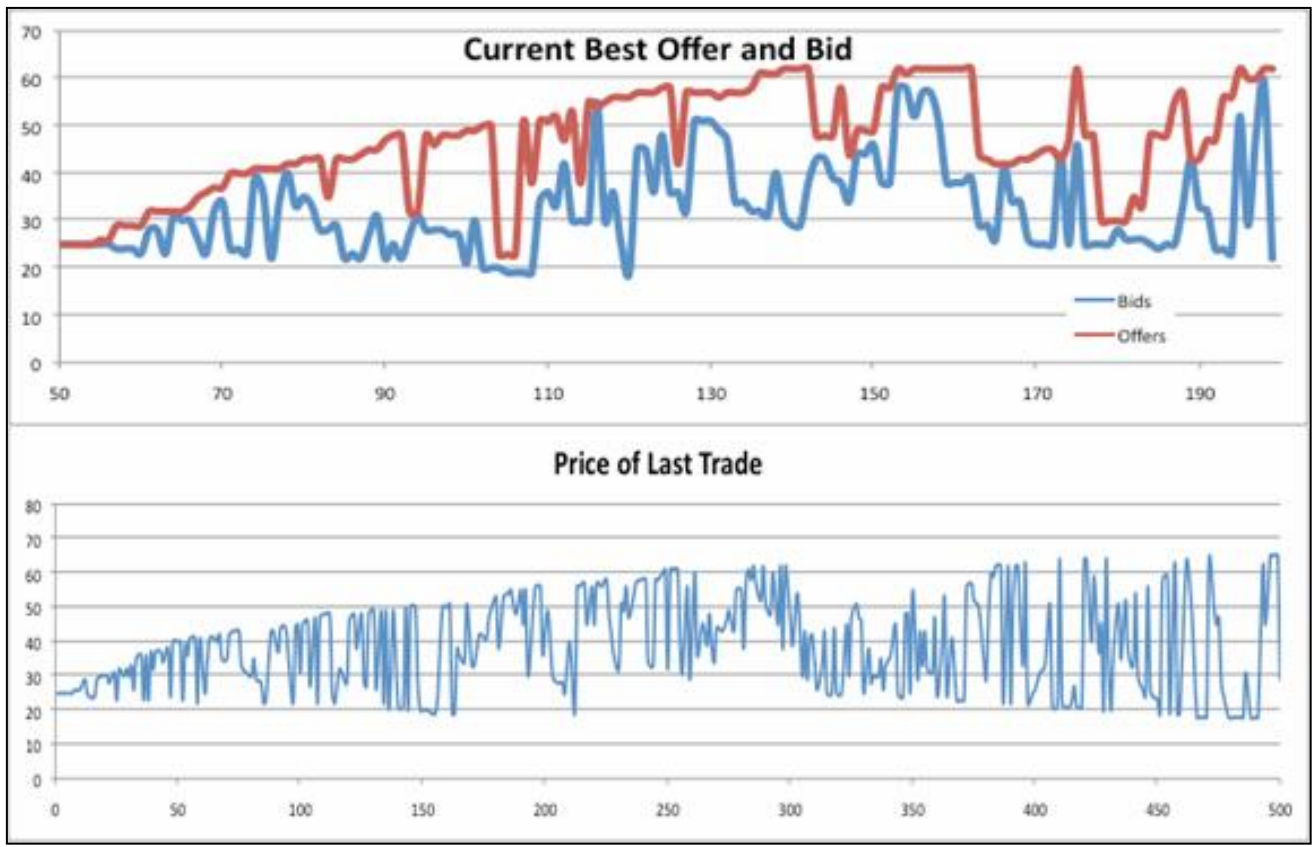

Fig 1. Spread Magnitude Dynamics and Price of Last Trade.

*2.3.1 The Zero Intelligence Model*

The Zero Intelligence Model (ZIM) is, essentially, an analytic (meaning data driven) model. It is a model of a single equity continuous double auction market; therefore, traders can both buy and sell and can do so at any time. There are two types of orders in the ZIM: market orders and limit orders. Market orders are orders that enter the market with an intent to buy or sell a certain number of shares and *do not* specify a particular price. Limit orders, on the other hand, enter the market with both a specified quantity of shares and a specified price. As market orders do not specify a price they are executed immediately upon entering the market at the best available price (see *infra*). Limit orders, however, will accumulate in the market until their specified price is met or they are cancelled. The accumulation takes place in a prioritized queue by price and arrival time. This accumulation of limit orders is called the order book. It is this accumulation of limit orders that creates liquidity (i.e., the ability for market orders to be executed) in the market. In the ZIM both market and limit orders arrive and are cancelled as a Poisson distributed process.

Even with these simple dynamics the behavior of the market model is reasonably similar to those of a real market. An important feature of these markets is the "spread", the distance in terms of price between the best offer to buy shares (i.e., the bid) of an asset and offer to sell an asset (i.e., the ask). The ask and the bid will change through time affecting the spread and the best available prices for market orders. This drift is caused by changes in the proportion of market orders to limit orders. If more market orders arrive than limit orders, the spread will increase as will volatility in the price. This will occur because an increase in market orders will begin to deplete the limit orders in the market. The price impact of a trade is another important feature of these systems. This simple model also appropriately predicts this feature, specifically, that the price impact function is concave [1]. This is caused by the increase in density of limit orders as market orders are executed against limit orders farther down the prioritized (i.e., first in, first out) queue.

*2.3.2 NetLogo and RepastS Implementations of the Zero Intelligence Model*

We instantiated the ZIM in NetLogo 4.0.2 and RepastS 1.0 so as to mitigate any potential nuances introduced by a particular modeling toolkit. The orders arrive at a Poisson generated rate. All orders are for one unit of the asset. Prices for orders are generated in one of two ways: 1) each order draws a unique price, or 2) a single price is drawn and then given to all arriving orders. The distribution for order price is a function of the current spread and consistent with the ZIM. While, technically the distribution for bid orders can range from $-\infty$ to the best offer and the distribution for offers ranges from the best bid to $+\infty$, within the simulations the distributions are bounded by the "market space" rather than infinity. As market orders arrive they are immediately executed against the best available limit order. At each time step existent limit orders have a random Poisson probability of being cancelled or expiring.

The simulations create dynamics that are consistent with the results reported by Iori et al [39] and Farmer et al. [38]. Both models create order book depth profiles that are sigmoidal in shape.

Furthermore, our simulations of the ZIM also demonstrate a random walk with respect to last trade price and the magnitude of the spread (i.e., distance between best Offer (upper line) and Bid (lower line), as shown in Figs 1 and 2 (Fig. 2 compares NetLogo and RepastS runs)[1]. The impact of a trade on price is also consistent between the simulations and the reported results (bottom graph in Fig. 1). Most trades have little impact. But the distribution is heavy-tailed, so while few trades have a large impact, the number of these high-impact trades cannot be ignored completely.

*2.3.3 The Heterogeneous Feedback Model*

The Farmer model concentrates on demonstrating market order-book structure and price behavior with respect to randomly placed (zero intelligence) trades, whereas the Cont Model [43 introduces the notions of heterogeneity and price feedback: Traders in the market are a heterogeneous group of agents with behaviors that are constantly modified in a feedback process with the market.

There are 5 behaviors exhibited by a wide range of markets and time periods:
1. Excess volatility
2. Heavy tails
3. Absence of autocorrelations in returns
4. Volatility clustering
5. Volume/volatility correlation

While, there are models that demonstrate these statistical properties, most are very complex making it difficult to determine where the statistical properties originate, leading to a diminished explanatory power of the model. This presents a particular problem for evaluation of the system in question and limits their utility for policy analysis and decision-support. The ZIM and Cont Model are very good examples of models being driven by a need for simplicity so the causes of dynamics and system properties are *highlighted* rather than obscured.

The Cont Model is a heterogeneous feedback model that describes a market where a single asset is traded by a set of agents. At each time step of the model, the agents all receive a common normally distributed "news" signal. Each agent compares that signal to its own unique threshold. If the signal exceeds the agent's threshold, an order is generated; otherwise the agent does not trade during that time period. The excess demand generated by all the agents' market orders causes the price to move, according to a linear price impact function.

The model is parsimonious, comprised of only 4 parameters: the frequency of updating agent thresholds, the standard deviation of the normally distributed "news" process, the market depth (affecting the slope of the price impact function), and the number of agents. Despite this simplicity the model has been shown to produce time series that capture the stylized statistical facts observed in asset returns.

*2.3.4 Implementation of Cont's Heterogeneous Feedback Model*

As with the Zero-Intelligence Model, our initial goal was to develop models in both NetLogo and RepastS, and then verify results consistent with those reported by Cont. This model was refined and implemented in RepastS to increase scalability and allow for the use of an extensive computational infrastructure (see [13] for more details on the Infrastructure for Complex-systems Engineering (ICE)).

---

[1] Fig. 2 also demonstrates good concordance between the Netlogo and RepastS models.

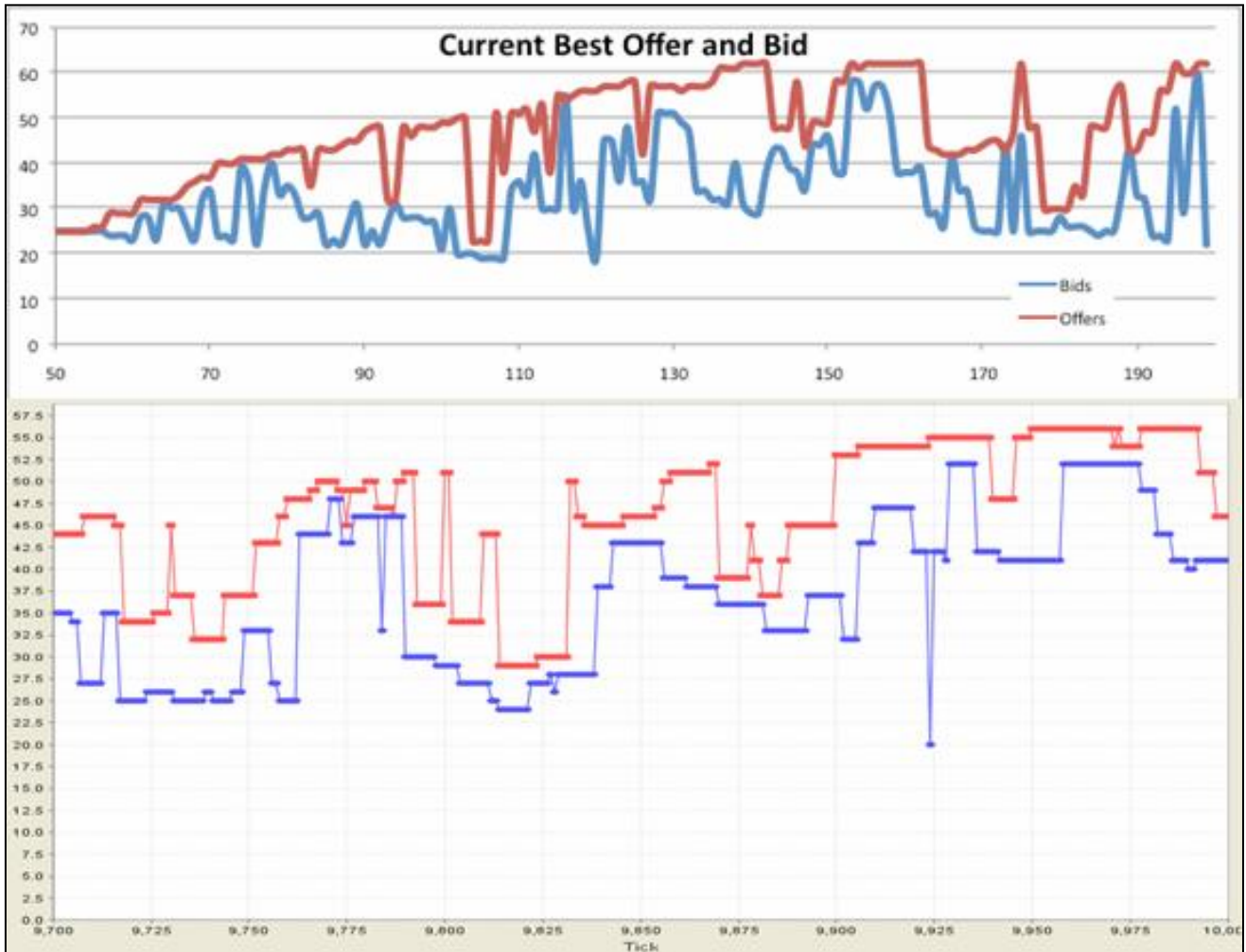

Fig 2. Spread Magnitude Dynamics between NetLogo and RepastS ZIMs.

*2.3.5 Docking and Relational Equivalence of the Models*

The next step for the simulation effort was to formally "dock" [4] the instantiations of the Farmer and Cont models together and relate them to the published results. For the purposes of this docking exercise, when possible, we will utilize a set of the "Cont statistics" as a common frame for comparison. Of note: the two models discussed herein (ZIM and Cont) compare their output to aggregated measures of system behavior (e.g., asset price, clustered volatility, and particular structures to the autocorrelation of returns); thus, placing them in Level 2 of Axtell's Empirical Relevance (i.e., macro-level quantitative correspondence with the real-world phenomena of interest) [14]. Due to the fact that agent-level correspondence is not at issue here, it is our contention that distributional equivalence (or a lack of statistical dissimilarity between distributions of results) is adequate for this docking exercise.

Unfortunately, we do not possess the original Cont or Farmer models or datasets from them; therefore, all that can be claimed is relational equivalence to the published results. Though not displayed here due to space limitations, the RepastS instantiation replicates the Cont results quite well but the results were less convincing for the NetLogo results. A likely explanation is that this seems to be a function of agent activation regimes (i.e. when agents trade and update).

Our implementation of the ZIM is generally consistent with those described in the Farmer works, of particular note is the random walk of the price, time series of best offer and bid, and sigmoidal shape of the order book. One may also produce Cont statistics for the Farmer model. When one does that it can be shown the Cont statistics for our RepastS and NetLogo instantiations are likely distributionally equivalent.

## 2.4 Integration of the Two Market Models

*2.4.1 Cont Market Model Extension*

A proof-of-concept extension was made of the Cont model to demonstrate that differentiating the information received by the trading agents would affect the returns received by those groups of agents. In the original Cont model agents received a random signal drawn from a normal distribution with a mean of zero. If the absolute value of this signal was greater than the agent's threshold the agent would sell if the signal was negative and buy if the signal was positive. Within the extension of the Cont model agents were divided into two groups: a large group that continued to receive the random signal and a small group that received information on the actual return being received lagged one time period. Fig. 3 shows the returns received by these two groups of agents (the red line is the small group of agents receiving the signal based upon actual returns). As can be seen in Fig. 3, the returns associated with the non-random signal deviate significantly from returns associated with the agents receiving the random signal. Fig. 4 shows the population returns and a 95% confidence interval around it. It should be noted that a heterogeneous Cont population produces slightly higher variance. It should be noted that Fig. 4 also shows a common feature of ABMs; namely, there is "burn in" at the beginning of this simulation. This means that the agents require a period of time after the simulation is started to "settle" down to the long run behavior. It will depend upon the application as the whether or not this is important for the analysis or if it can be discarded.

*2.4.2 Integration of Cont's Traders with Farmer's Market Structure*

An extended model was implemented, combining Cont heterogeneous trading agents in a Farmer order-book market. Inserting heterogeneous trading agents into the order-book based market model required three extensions:

- Random trades (*a la* Farmer) were eliminated by inserting trading agents from the Cont trading model.
- Trades generated by the Cont agents were executed using limit orders from the order book. The order book was otherwise populated with random limit order arrivals, as in the original Farmer model.
- Actual price returns were used in place of the price impact function since actual transaction information is available from the order book.

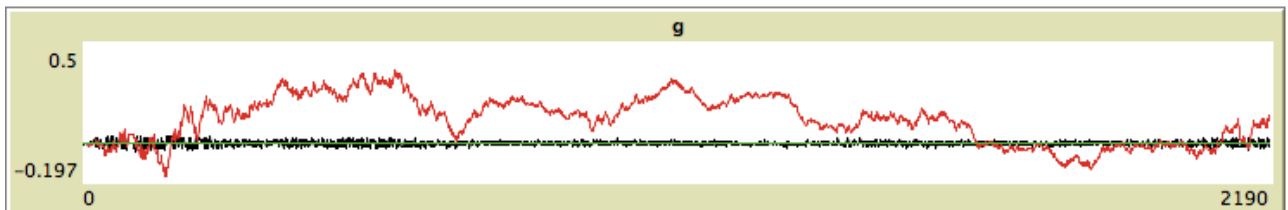

Fig 3. Differentiated Returns for a Small Group of Agents Receiving Additional Information.

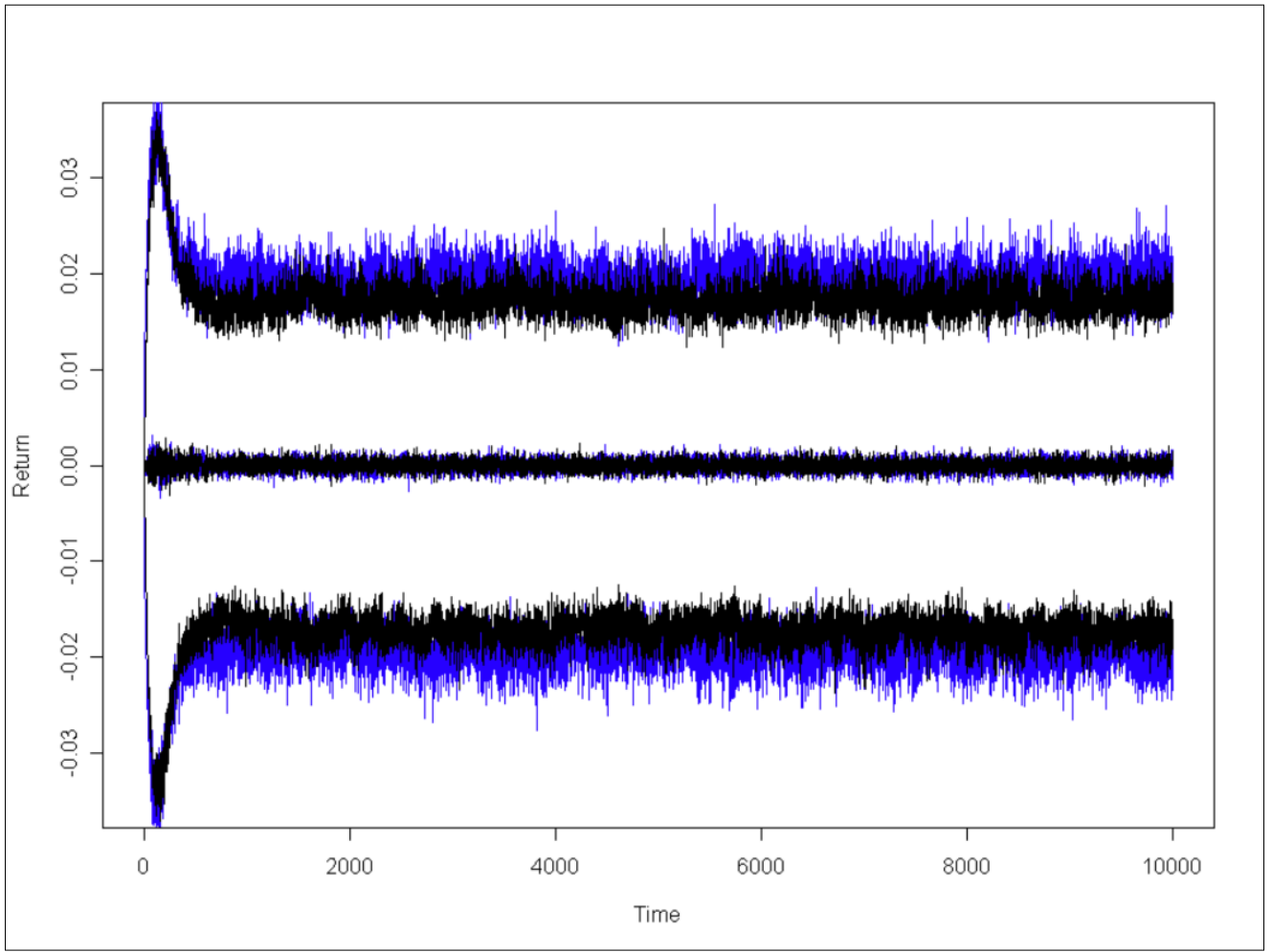

Fig 4. The 95% Confidence Interval for Standard Cont and Extended Cont.

The numerical results for the extended model are very similar to the original zero-intelligence Farmer model. This dynamic lends support to the idea that the structure of the market has a great deal of impact on the functioning of the market.

### 2.5 Including Margin

Furthering the idea that market structure is a primary determinant of market dynamics, Thurner and colleagues [15] employ a market model without an order book that includes noise traders and funds able to use leverage to increase long positions in response to a market price lower than a commonly perceived intrinsic share value. They show that higher leverage limits on these funds increase kurtosis on the distribution of returns and generate clustered volatility. Importantly from the perspective of a market regulator, they also show that individual margin requirement--though rational from the individual perspective—can result in a coordinated margin call that induces market price crash.

Incorporating the idea of leveraged value funds into the confines of an order book, the MITRE team modified the fund models to participate in a continuous double auction. The inclusion of such funds into the simulated markets defined in previous sections still results in price time series that match expected forms. Work is still underway to define the market's susceptibility to price crash given parameter inputs, and to dock the order-book based market with previous results.

### 3. Model Validation

As noted above, the MITRE team demonstrated relational equivalence between its implementation of an agent-based market model and two then-current models of financial markets: Farmer's order-book based market model using Zero-Intelligence (ZI) traders and Cont's market model using heterogeneous traders. The MITRE work continued, by combining the Cont and Farmer modes and by extending them. In this subsequent work, our modeling efforts have concentrated on adding a higher level of verisimilitude to the models, initially by adding heterogeneous traders to the order-book model, and more recently by developing an order-book based market incorporating hedge-fund traders.

In aligning these extensions with the work of Cont and Farmer, we have also followed their lead in validating our market models. The prior literature [1, 2, and 16], discusses and demonstrates model alignment with a number of stylized facts which have been shown to occur empirically in financial markets. As market models become more sophisticated, it becomes more difficult to validate their behavior by purely analytic arguments based on their mechanics. In this work, we adopt three key stylized facts:

i. *Heavy tails.* The distribution of daily and hourly returns displays a heavy tail with positive excess kurtosis.
ii. *Absence of autocorrelations in returns.* Autocorrelations of asset returns are often insignificant, except for very small intraday timescales)
iii. *Volatility clustering.* While returns themselves are uncorrelated, absolute returns $|rt(\_)|$ display a positive, significant and slowly decaying autocorrelation function: $corr(|rt|, |rt+\_|) > 0$ for _ ranging from a few minutes to several weeks.

Figure 5 follows the format set forth in [3] and displays data for the Standard and Poor's 500 index of the large-capitalization U.S. equities market, spanning 1980-Sept. 2010. Notice how the Distribution of Returns demonstrates heavy tails (i.e. a kurtosis greater in magnitude than that of a normal distribution) as described in (i) above. For the autocorrelations of returns, both inspection and quantification lead to a lack of correlation conclusion (fact ii above). Finally, the autocorrelation of the absolute value of returns exhibits a positive, but slowly decaying value, as expected in fact iii above.

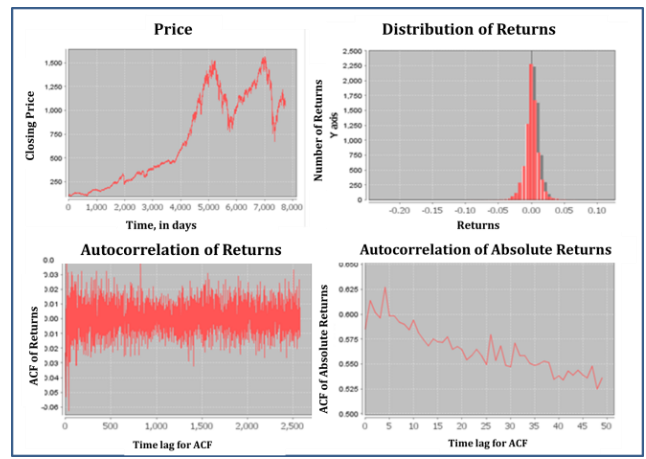

Figure 5 - Stylized facts, computed for the S&P500 index spanning 1980-September 2010.

Figure 6 depicts the same computations for the execution of our most recent implementation of the leveraged value fund model with an order book.

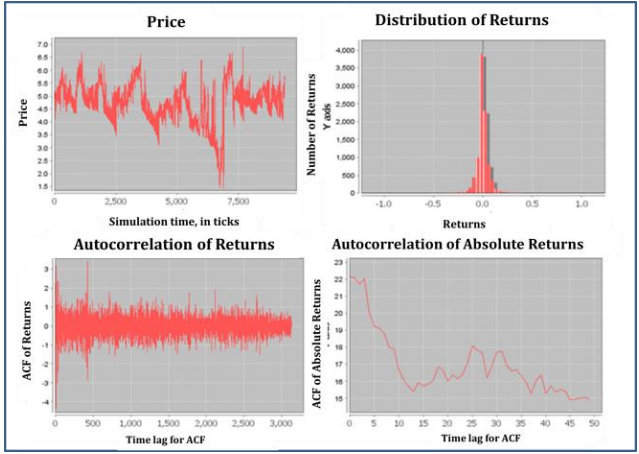

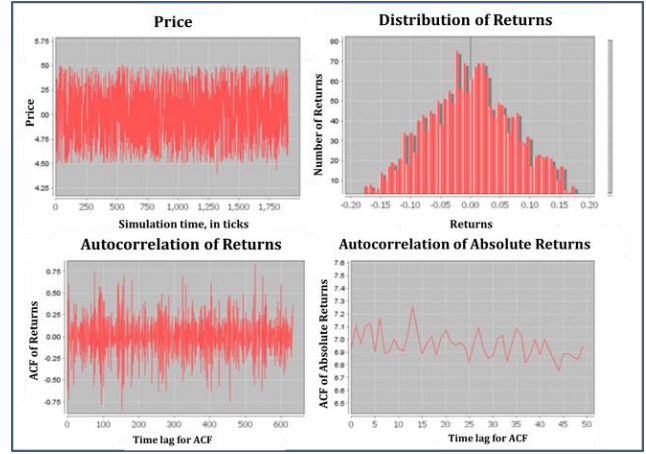

Figure 6 - Stylized facts, computed for the execution of MITRE's hedge fund market model.

Figure 7 -- Stylized facts--degenerate situation--computed for the execution of MITRE's hedge fund market model. In this case, the kurtosis of the distribution of returns is very close to that of a normal distribution. Not also the decidedly white-noise appearance of the underlying price time series (thin trading).

Again, the three display the characteristics observed in empirical markets.

Note that we don't expect an exact match in the graphs from any model and the empirical markets. Different model parameter settings and different stochastic conditions will lead to different details in the price and return time series. We are merely trying to duplicate the general features of returns. To make this clear, compare the similarity of the key graphs in Figure 5 and Figure 6 to the similarities between those in Figure 5 and Figure 7. Figure 7 depicts a "degenerate" distribution, where agents in the simulation were limited by the number of shares they could trade. The two autocorrelation functions appear to adhere to facts (ii) and (iii), but the fitness is penalized by lack of fat tails.

We have further extended the use of these stylized facts as a fitness function for our models. For each fact, we have established a quantitative measure we combine in a linear sub to compute model fitness as a function of the extent to which a model execution's computed statistics correspond with the expectation. Kurtosis is used to measure (i) Heavy Tails. Correlations coefficients are used to measure the (ii) Autocorrelations of returns and (iii) Volatility Clustering.

Figure 8 depicts the linear summation fitness computations for a full-factorial experiment (parameter sweep), used as a validation exercise in our model development. During this particular experiment, 2,200 different parameter sets were explored. executions of our model in a subset of our parameter space. For each a market model is run as described above, and the fitness value is computed as a weighted sum of the stylized fact computations described above. Low fitness computations correspond with parameter settings that will be discarded because those settings result in a market which fails to meet one or more of the stylized statistical properties. For example, for the model that resulted in Figure 7, the low fitness resulted from a distribution of returns with low kurtosis (NB: this is run 1378 in Figures 8 and 9).

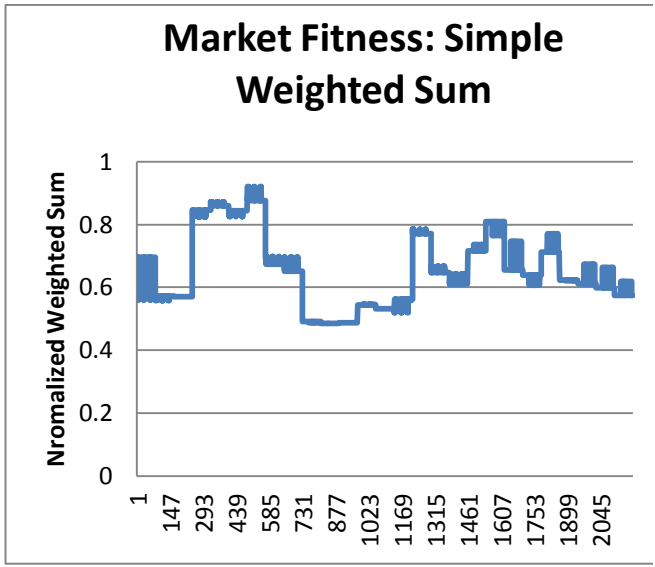

Figure 8 -- Market Fitness for the model, as a sum of statistical compliances with stylized facts, for a sweep over 2200 unique parameter sets.

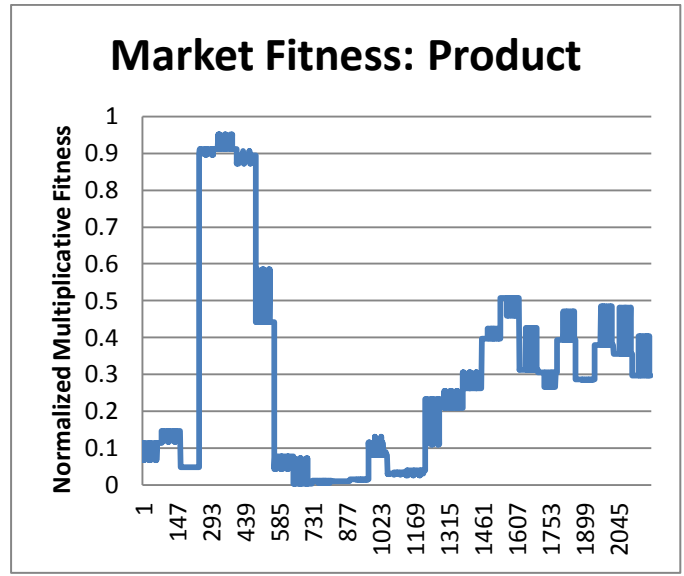

Figure 9 -- Market Fitness for the model, as a product of statistical compliances with stylized facts, for a sweep over 2200 unique parameter sets.

Figure 9 examines the fitness of the same set of runs, using the product of the three observations in lieu of the summation. Note that the product-based fitness has a broader range. This is because in cases where one or more statistics fail to correspond to expectation (e.g., returns are not leptokurtotic, as in Figure 7), the summation allows for the remaining compliant statistics (e.g., there is no autocorrelation of returns in Figure 7) to deceptively counterbalance the offending statistic. In contrast, the product offsets this counterbalancing force by scaling the fitness down in cases where one or more stylized facts are not observed.. Since highest fitness values should go to those models which meet all three facts, this scaling of fitness—or penalty—is useful in cases such as that found in Figure 3, where a penalty is needed for low kurtosis.

Future work in this area includes the further development of these fitness functions, with particular emphasis on broadening our list of stylized facts (many of which are satisfied by our model but which are not shown here), and selecting the best scoring function for combining scores from individual facts. In addition, it is worth exploring the application of the fitness functions beyond validation, employing it as an optimization tool for selecting model parameters.

### 4. Fractal and Multi-Fractal Analysis

In addition to the Cont stylized facts about markets, we consider Mandelbrot's claims of the fractal and multi-fractal nature of financial time series (see generally, Mandelbrot [17]) and began an effort to incorporate fractal and multi-fractal analysis tools in our research. Most individuals are familiar with geometric fractal figures such as the Sierpinski Gasket and the Mandelbrot Set, however time series data can also have a fractal structure. Furthermore, if the mechanism that generated the time series changes over time or over scale then the time series will have a multi-fractal structure [18]. We have implemented a number of methods to calculate the fractal and multi-fractal dimensions of a time series. The present discussion will focus on the well known "box counting" algorithm [19].

Figure 10a shows the time series data used in this analysis. The y-axis is daily close and the x-axis is time units. Time units for the S&P 500 are days. The time unit for the simulated data is an abstract time step. In all cases 6546 time units (days or steps) were collected for analysis. The top time series is from our simulated equity market. The middle time series is S&P 500 data from 1/3/1950 to 2/12/1976. Finally, the bottom time series is S&P 500 data from 11/21/1984 to 11/3/2010. As can be seen, the simulated time series is the smoothest with "old" S&P 500 being less smooth and the "new" S&P 500 data being the least smooth. Also of note is the scale, approximately two orders of magnitude separate the time series (as can be seen in Figure 10b). All of these differences make meaningful comparisons difficult. In these situations a fractal analysis can be very enlightening. The fractal perspective is one of structure rather than value. This perspective lets one compare the fine structure of a time series and begin to make statements about the mechanism that generated it.

Figure 11 shows the basic components of our multi-fractal analyses of the aforementioned time series. Figure 11a shows the scaling exponent of the three time series. There are two points to take away from Figure 11a: 1) The scaling exponent does not change linearly, there is a dogleg in the line. This means that there is a multi-fractal process at work, and 2) the simulation time series and the "old" S&P 500 time series are almost identical. Now that we know a multi-fractal process is at work, we turn to Figure 11b, a chart of the multi-fractal spectra of the data sets. As can be seen in Figure 11b, the time series from the simulation and the time series from the "old" S&P 500 are very similar, with the "new" S&P 500's spectrum being quite different. This provides evidence that the dynamics captured by the simulation are a reasonable representation of the dynamics seen in the S&P 500 between 1950 and 1970, *despite the order of magnitude difference seen in the value of the close*. However, the simulation does not adequately capture the dynamics of the S&P 500 as it exists today. Additional analyses are ongoing to determine when the multi-fractal spectra diverge and what may have caused the divergence.

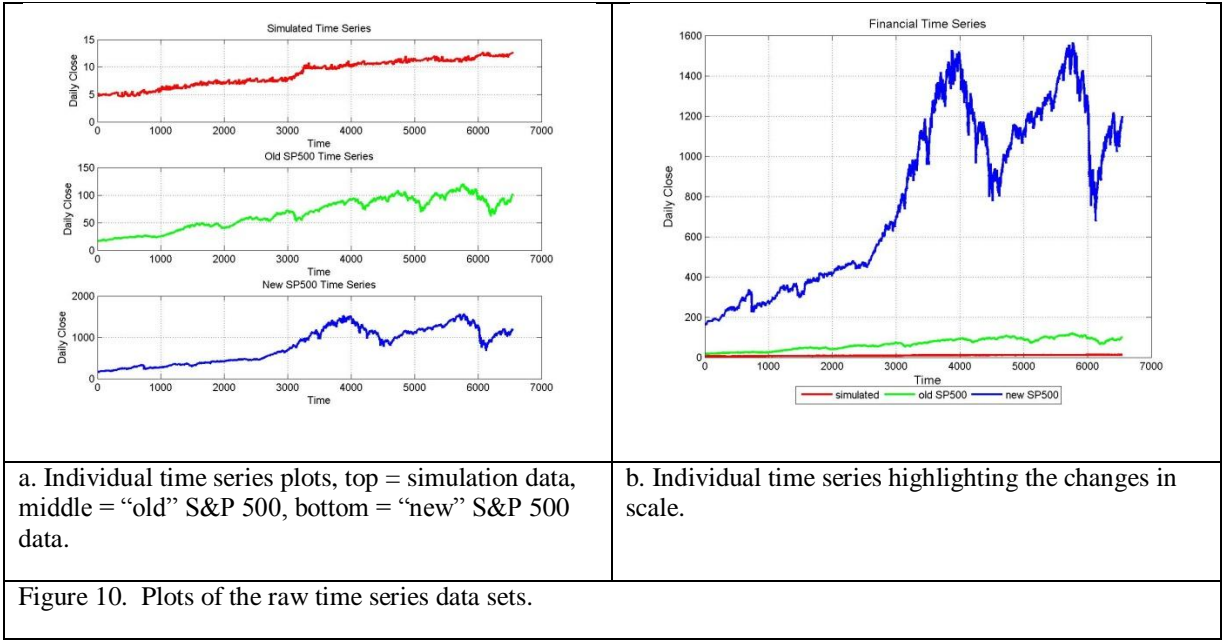

a. Individual time series plots, top = simulation data, middle = "old" S&P 500, bottom = "new" S&P 500 data.

b. Individual time series highlighting the changes in scale.

Figure 10. Plots of the raw time series data sets.

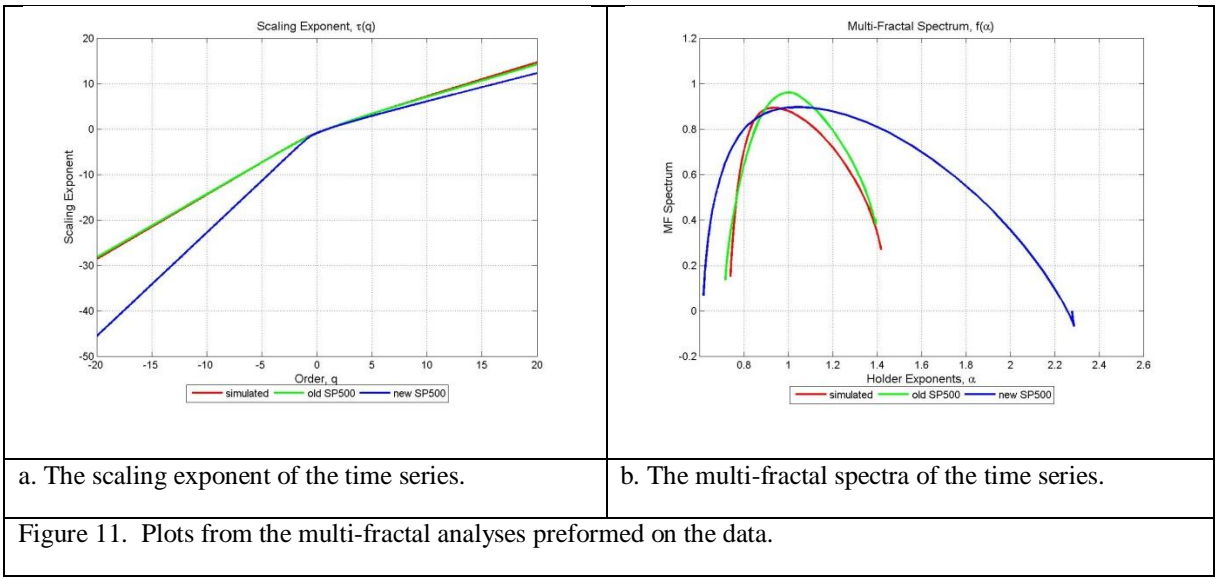

a. The scaling exponent of the time series.

b. The multi-fractal spectra of the time series.

Figure 11. Plots from the multi-fractal analyses preformed on the data.

## 5. CONCLUSION

As our economy becomes increasingly interconnected with other critical infrastructural components of our nation, our ability to regulate it and understanding the impact of changes to the regulation of it will continue to increase in importance. As financial markets are systems made up of many interacting components and the system has many feedbacks, the most efficient way to understand these systems is to simulate them [20]. The work discussed herein represents our initial steps in creating a policy analysis/decision-support system to foster a deeper understanding.